\documentclass[12pt]{article}
\usepackage{graphicx} 
\usepackage{amsmath}
\usepackage{mathtools}
\usepackage{here} 
\usepackage{hyperref}
\hypersetup{
setpagesize = false,
bookmarksnumbered = true,
bookmarksopen = true,
colorlinks = true,
linkcolor = blue,
citecolor = blue,
}
\usepackage[hang,small,bf]{caption} 
\usepackage[subrefformat=parens]{subcaption}
\usepackage[margin=30truemm]{geometry} 
\numberwithin{equation}{section} 
\usepackage{cite} 
\usepackage{amssymb} 

\title{
\begin{flushright}
\ \\*[-80pt]
\begin{minipage}{0.2\linewidth}
\normalsize
HUPD-2301 \\*[50pt]
\end{minipage}
\end{flushright}
{\Large \bf
Approach to the arbitrariness problem of boundary conditions in $S^1/Z_2$ brane-world models
\\*[20pt]}
}

\author{\centerline{
Kota Takeuchi$\,^{1}$\footnote{k-takeuchi@hiroshima-u.ac.jp}$\,\,$
Tomohiro Inagaki$\,^{1,2,3}\,$\footnote{inagaki@hiroshima-u.ac.jp
}}\\*[20pt]
\centerline{
\begin{minipage}{\linewidth}
\begin{center}
$^1${\it \normalsize
Graduate School of Advanced Science and Engineering, Hiroshima University,
Higashi-Hiroshima~739-8526,~Japan \\*[5pt]
$^2${\it \normalsize
Information Media Center, Hiroshima University, Higashi-Hiroshima 739-8521, Japan} \\
$^3${Core of Research for the Energetic Universe, Hiroshima University, Higashi-Hiroshima 739-8526, Japan}
}
\end{center}
\end{minipage}}
\\*[50pt]}
\date{}

\begin{document}
\maketitle

\begin{abstract}
We study the arbitrariness of boundary conditions (BCs) on $S^1/Z_2$ brane-world models with the gauge group $U(N)$.
The BCs are chosen independently on the branes and the bulk in this model.
There are numerous choices for BCs in general, but some of the BCs are connected through gauge transformations.
We show that the equivalent relations are obtained on the UV-brane without relying on specific transformation parameters.
There is no other equivalent relation on the UV-brane.
On the other hand, we find that a gauge transformation with a kink connects all the BCs on the bulk and IR-brane.
It means that the arbitrariness of BCs is completely solved on the bulk and the IR-brane except for the UV-brane.
\end{abstract}


\section{Introduction} \label{sec1}
In Gauge Higgs Unification (GHU) theories, the Higgs bosons are introduced as extra-dimensional components of the gauge fields, that is the Higgs bosons are unified with the gauge bosons. 
Especially, the GHU theories with an orbifold space are phenomenologically attractive.
It has been shown that 4D chiral fermions are naturally generated, and the Higgs mass splitting can be elegantly realized \cite{MANTON1979141, FAIRLIE197997, DBFairlie_1979, 10.1143/PTP.103.613, 10.1143/PTP.105.999, PhysRevD.64.055003, doi:10.1142/S0217732302008988, SCRUCCA2003128, PhysRevD.69.055006}.
Various models with extra-dimensions can be constructed by combining a variety of contents such as the structure of space-time, field contents, gauge symmetry, geometric symmetry, and boundary conditions (BCs) imposed on fields.

Realistic combinations of the first four have been actively studied \cite{BURDMAN20033, PhysRevD.70.015010, PhysRevD.67.085012, AGASHE2005165, PhysRevD.78.096002, PhysRevD.79.079902, 10.1093/ptep/ptu146, PhysRevD.104.115018, HOSOTANI2005276, PANICO2006186, PANICO2007189, 10.1093/ptep/ptz083, PhysRevD.98.015022, PhysRevD.106.055033}.
For the last one, however, there remains the arbitrariness problem of BCs.
The definition of a model with compactified space requires BCs. 
There are numerous choices for BCs and the choices lead to physically different models.
We are unknown which type of BCs should be selected without relying on phenomenological information.
This is the arbitrariness problem of BCs\cite{doi:10.1142/5326, Quiros:2003gg, HABA2003169}.

This problem has been partly eliminated.
Ref.\cite{HABA2003169} pointed out that some BCs are connected by gauge transformations.
The connected BCs produce a physically equivalent model.%
\footnote{
The physical equivalence of the BCs in one EC is guaranteed by the Aharonov-Bohm (AB) phase based on the Hosotani mechanism\cite{HOSOTANI1983309, HOSOTANI1983193, HOSOTANI1989233}.}
We define equivalence classes (ECs), each of which consists of BCs connected by gauge transformations.
There are multiple ECs in an orbifold model in general and classifying ECs is the first step to solving the arbitrariness problem of BCs.
Many works have been done to study ECs in various orbifold models by using specific gauge transformations\cite{10.1143/PTP.111.265, PhysRevD.69.125014, 10.1143/PTP.120.815, 10.1143/PTP.122.847, doi:10.1142/S0217751X20502061, Kawamura2023}.
Our previous paper\cite{10.1093/ptep/ptae027} has achieved the sufficient classification of ECs without relying on an explicit form of gauge transformations.
As a result, the number of ECs in $U(N)$ and $SU(N)$ gauge theories on $S^1/Z_2$ orbifold is precisely obtained to be $(N+1)^2$ \cite{10.1143/PTP.111.265}.

\begin{figure}[ht]
\centering\includegraphics*{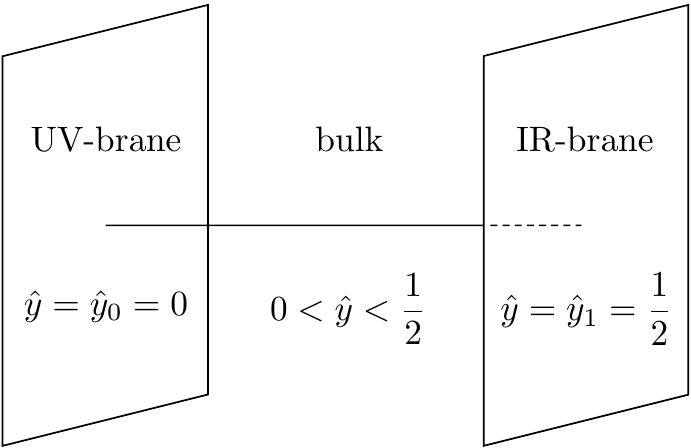}
\caption{$S^1/Z_2$ brane-world}
\label{FigBrane}
\end{figure}

In this paper, we consider $S^1/Z_2$ brane-world model, where the BCs can be chosen independently on the branes and the bulk (see Fig.\ref{FigBrane}).
As a result, the BCs cannot be connected by constant parameters, but the equivalent relations are obtained on the UV-brane by coordinate-dependent parameters.
It is a sufficient classification because this study does not rely on an explicit form of gauge transformations.
On the other hand, we also show that all the BCs on the bulk and the IR-brane can be connected by a gauge transformation with a kink.
It means that the arbitrariness of BCs is completely solved on the bulk and IR-brane except for the UV-brane.

In Section \ref{sec2}, the gauge transformations of BCs on $S^1/Z_2$ are discussed.
In Section \ref{sec3}, we consider the case that the transformation parameters are constant.
In Section \ref{sec4}, the connections between the BCs are investigated separately on the UV-brane and the others.
Section \ref{sec5} presents the conclusion.

\section{Boundary Conditions and Equivalence Classes} \label{sec2}
In this paper, we focus on $U(N)$ gauge theory on a four-dimensional Minkowski space-time $M^4$ and one extra-dimensional space compactified on the $S^1/Z_2$ orbifold, which is identified as two points on a circle $S^1$ with radius $R$ by parity.
Let $x^\mu \,(\mu=0,1,2,3)$ and $y$ be the coordinates of $M^4$ and $S^1/Z_2$, respectively.
(Hereafter the subscript $\mu$ is omitted). 
5D fermion fields $\Psi(x,y)$ and 5D gauge fields $A_M(x,y)$ respectively belong to $U(N)$ fundamental and adjoint representations. 
The action is given by
\begin{equation}
    S = \int d^4 x  dy \, \mathcal{L}^{5D} 
    =\int d^4 x  dy \, \left\{ -\frac{1}{4}F_{MN}F^{MN} +  \bar{\Psi} i \Gamma^M D_M  \Psi \right\},
\end{equation}
where $x^M=(x^\mu, y)$ indicate 5D coordinates, and $\Gamma^M = (\gamma^{\mu},i\gamma^5)$ are Dirac gamma matrices ($\gamma^5 \equiv i \gamma^0 \gamma^1 \gamma^2 \gamma^3$).
The covariant derivative and the field-strength-tensor are denoted as, $D_M = \partial_M + ig A_M(x,y)$ and $F_{MN} = \frac{i}{g} \left[ D_M, D_N \right]$, respectively.
We identify any point on the y-coordinate as, $y \sim y+2\pi R \sim -y$, and the fundamental region can be written as, $0 \leq y \leq \pi R$.

The 5D Lagrangian on $M^4\times S^1/Z_2$ is invariant under the following transformations:
\begin{equation} \label{trans}
    \mathcal{\hat{T}}: y \to y+ 2 \pi R ,\,\,\,\,\,\,\,\,\,
    \mathcal{\hat{P}}_0 : -y \to +y ,\,\,\,\,\,\,\,\,\,
    \mathcal{\hat{P}}_1 : \pi R -y \to \pi R +y.
\end{equation} 
$\mathcal{\hat{T}}$ is a translation along the circle. 
$\mathcal{\hat{P}}_0$ and $\mathcal{\hat{P}}_1$ are respectively parity transformations around the two fixed points, $y=y_0=0$ and $y=y_1=\pi R$, and they satisfy, $\mathcal{\hat{P}}^2_0 = \mathcal{\hat{P}}^2_1 = \mathcal{\hat{I}}$, where $\mathcal{\hat{I}}$ is an identity transformation. 
We note that only two of them are independent because of $\mathcal{\hat{T}}= \mathcal{\hat{P}}_1\mathcal{\hat{P}}_0$ is satisfied.
We take $(P_0,P_1)$ as the basic boundary conditions (BCs).
The BCs of each field are represented as
\begin{align}
    \Psi(x, y_i -y) &= P_i \gamma^5 \Psi(x, y_i +y), \\
    A_{\mu}(x, y_i -y) &= P_i A_{\mu}(x, y_i +y) P_i, \\ 
    A_{y}(x, y_i -y) &= -P_i A_{y}(x, y_i +y) P_i ,
\end{align}
where $N\times N$ representation matrices $P_i \,(i=0,1)$ satisfy parity and unitary conditions, $P_i=P_i^{-1}=P_i^{\dag}$.
These are Hermitian $U(N)$ matrices.

There are numerous choices for the BCs because any $(P_0,P_1)$ is allowed as long as they satisfy parity and unitary conditions.
The choice of them lead to physically different models\cite{doi:10.1142/S0217732302008988}.
However, several BCs are connected by gauge transformations\cite{HABA2003169}.
Let us transform each field with a transformation matrices $\Omega(x,y)$.
The BCs of the transformed fields are written as
\begin{align}
    \Psi'(x, y_i -y) &= P'_i \gamma^5 \Psi'(x, y_i +y),\\
    A'_{\mu}(x, y_i -y) &= P'_i A'_{\mu}(x, y_i +y) P_i^{'\dag}
    - P'_i  \, \partial_\mu  \, P_i^{'\dag},\\
    A'_y(x, y_i -y) &= -P'_i A'_y(x, y_i +y) P_i^{'\dag}
    - P'_i \left( -\partial_y \right) P_i^{'\dag},
\end{align}
where
\begin{align} 
    P'_i=\Omega(x, y_i-y) P_i \Omega^\dag(x, y_i+y). \label{P'}
\end{align} 
The transformed set $(P'_0,P'_1)$ generally depend on the coordinates, but if it remains constant and satisfies parity and unitary conditions, then $(P'_0,P'_1)$ can be another choice for $(P_0,P_1)$.
In this case, $(P_0,P_1)$ and $(P'_0,P'_1)$ are generally different but yield a physically equivalent model, denoted as $(P_0, P_1) \sim (P'_0, P'_1)$.
Such connected sets of BCs belong to the same equivalence class (EC), each of which consists of physically equivalent BCs.
BCs-connecting gauge transformation is defined by the transformation satisfying the following conditions:
\begin{equation} \label{ECsgauge}
    \partial_M P'_0 = 0, \,\,\,\,\, 
    \partial_M P'_1 = 0, \,\,\,\,\,
    P'^{\dag}_0 = P_0, \,\,\,\,\,
    P'^{\dag}_1 = P_1. 
\end{equation}
Both $(P_0,P_1)$ and $(P'_0,P'_1)$ remain constant and satisfy parity and unitary conditions under this transformation.

How is each EC characterized?
$P_0$ and $P_1$ are not generally diagonal, but Ref.\cite{10.1143/PTP.111.265} have shown that they can be diagonalized simultaneously by gauge transformations.
It means that each EC contains at least one diagonal set, which consists of a diagonal $P_0$ and a diagonal $P_1$. 
All ECs on $S^1/Z_2$ are characterized the sets of $(P_0,P_1)$'s eigenvalues.
A diagonal set $(P_0, P_1)$ is rearranged to
\begin{align}
\begin{split}
    P_0 &= \text{diag} \overbrace{ (  +1, \dotsc, +1, +1, \dotsc, +1,-1, \dotsc, -1,-1, \dotsc, -1  )}^N,\\
    P_1 &= \text{diag} ( \underbrace{ +1, \dotsc, +1 }_p , \underbrace{ -1, \dotsc, -1 }_q , \underbrace{ +1, \dotsc, +1 }_r , \underbrace{ -1, \dotsc, -1 }_{s=N-p-q-r} ),
\end{split}
\end{align}
where $p,q,r,s$ denote non-negative integers and satisfy, $0 \leq p,q,r,s \leq N $ and $p+q+r+s=N$. 
A diagonal set $(P_0,P_1)$ is characterized by a set of non-negative integers:
\begin{equation}
    [p,q,r,s].
\end{equation}
Each EC contains s diagonal set, but there remains a possibility that it contains more than one, so that we need to investigate whether one diagonal set is transformed into another diagonal set under gauge transformations.
It has been investigated by using specific gauge transformations in previous paper\cite{10.1143/PTP.111.265}.
In this paper, we examine it by using the general gauge transformations with arbitrary parameters.
The classification of ECs is completed after this sufficient investigation.


\section{Analysis of global gauge transformations} \label{sec3}

In this section, we discuss the case that the transformation matrices $\Omega$ are coordinate-independent, called global gauge transformations.
It is shown that the global gauge transformations do not achieve the connection between the diagonal sets of BCs.
This discussion will be useful for the analysis of the local gauge transformations in Section \ref{sec4}.

A transformation matrix is generally represented as $\Omega (y) = \exp{[i f^a (\hat{y}) T^a ]}$.
Here $\hat{y} \equiv y / 2\pi R$ is a dimensionless parameter, $f^a (\hat{y})$ is a continuous real parameter, and $T^a$ is a generator of the $U(N)$ group $(a=1, \ldots, N^2)$.
We drop $x$-dependence since it does not contribute to our discussion.
In order to investigate the connection between the diagonal sets of BCs, the generators $\{ T^a (a=1, \ldots, N^2) \}$ are classified into the following four types \cite{doi:10.1142/S0217751X20502061}:
\begin{align}
    &\text{(i) the commutative type:}\quad\quad\quad\quad
    T^{a_{++}} =
    \begin{pmatrix}
    \star & 0 & 0 & 0 \\
    0 & \star & 0 & 0 \\
    0 & 0 & \star & 0 \\
    0 & 0 & 0 & \star 
    \end{pmatrix}, \label{gene_com} \\[6pt]
    &\text{(ii) the mixed type:}\quad\quad
    T^{a_{+-}} =
    \begin{pmatrix}
    0 & \star & 0 & 0 \\
    \star & 0 & 0 & 0 \\
    0 & 0 & 0 & \star \\
    0 & 0 & \star & 0 
    \end{pmatrix} ,\quad
    T^{a_{-+}} =
    \begin{pmatrix}
    0 & 0 & \star & 0 \\
    0 & 0 & 0 & \star \\
    \star & 0 & 0 & 0 \\
    0 & \star & 0 & 0 
    \end{pmatrix}, \label{gene_mix} \\[6pt]
    &\text{(iii) the anti-commutative type:}\quad
    T^{a_{--}} =
    \begin{pmatrix}
    0 & 0 & 0 & \star \\
    0 & 0 & \star & 0 \\
    0 & \star & 0 & 0 \\
    \star & 0 & 0 & 0 
    \end{pmatrix}, \label{gene_anti}
\end{align}
where $\star$ stands for non-zero sub-matrices and $0$ is a sub-matrix with all components zero (a null sub-matrix).%
\footnote{The concrete forms of the generators are different in some cases, such as $p=0$ and $N=2,3$.
The following discussion in this section is also applied to such cases.}
\footnote{There are gauge transformations with multiple types of the generators, but they cannot construct a unique connection between the BCs in many cases.}
The numbers of $T^{a_{++}}$ , $T^{a_{+-}}$ , $T^{a_{-+}}$ and $T^{a_{--}}$ are respectively $p^2+q^2+r^2+s^2$, $2(pq+rs)$, $2(pr+qs)$ and $2(ps+qr)$, and the total number of them is $N^2$. These generators commute or anti-commute with $P_0$ and $P_1$:
\begin{gather}
    \left[ T^{a_{++}}, P_0 \right] = \left[ T^{a_{++}}, P_1 \right] = 0, \\
    \left[ T^{a_{+-}}, P_0 \right] = \left\{ T^{a_{+-}}, P_1 \right\} = 0, \\
    \left\{ T^{a_{-+}}, P_0 \right\} = \left[ T^{a_{-+}}, P_1 \right] = 0, \\
    \left\{ T^{a_{--}}, P_0 \right\} = \left\{ T^{a_{--}}, P_1 \right\} = 0, 
\end{gather}
where $[A,B] \equiv AB-BA$ and $\{A,B\} \equiv AB+BA$.

We consider global gauge transformations $\Omega_{\pm}$.
They have the constant parameters, $f^{a_{\pm}} = c^{a_{\pm}}/2$ $\,(c^{a_{\pm}} \in \mathbb{R})$, and the generators $T^{a_{\pm}}$, respectively.
Let us transform a matrix $P$ by $\Omega_{\pm}$, where $T^{a_{+}}$ commutes with $P$ and $T^{a_{-}}$ anti-commutes with $P$.
The transformation of $P$ (\ref{P'}) is calculated as
\begin{align}
    &P' = \Omega_+ P \, \Omega_+^{\dag} = \Omega_+ \Omega_+^{\dag} P = P ,  \label{GL+} \\ 
    &P' = \Omega_- P \, \Omega_-^{\dag} = \Omega_- \Omega_- P = e^{i T^-}P , \label{GL-}
\end{align} 
where $T^- \equiv c^{a_-} T^{a_-}$ is a Hermitian matrix.
Hereafter we employ the notation: $T^{st} \equiv c^{a_{st}} T^{a_{st}}$ $(s,t = + \, \text{or} \, -)$.
(\ref{GL+}) and (\ref{GL-}) lead to the following results under the global gauge transformations:
\begin{itemize}
    \item[(i)$\,\,$] 
    The commutative type only becomes the identity transformation: \\
    $(P'_0, P'_1) = (P_0, P_1)$.
    \item[(ii)$\,$] 
    The mixed type rotates one matrix with the other matrix fixed: \\
    $(P'_0, P'_1) = (P_0, e^{i T^{+-}} P_1)$ or $(P'_0, P'_1) = (e^{i T^{-+}} P_0, P_1)$.
    \item[(iii)] 
    The anti-commutative type rotates both simultaneously: \\
    $(P'_0, P'_1) = (e^{i T^{--}} P_0, e^{i T^{--}} P_1)$.
\end{itemize}
Is a diagonal set $[p,q,r,s]$ transformed to a different diagonal set $[p',q',r',s']$ through such rotations?
The answer is NO.
The proof will be done step-by-step for the cases that the generators are $2\times 2$, $N\times N$ with two blocks, and $N\times N$ with four blocks.

\subsection{$2\times 2$ matrices} \label{sec3.1}
The first step is to consider the following $2\times 2$ generator:
 \begin{equation}
    T^- =
    \begin{pmatrix}
    0 & c \\
    c^* & 0 
    \end{pmatrix},
\end{equation}
where $c$ is a complex number.
This generator produces the mixed and anti-commutative type transformations in $U(2)$ sub-group in $U(N)$.
The $n$-th term in the Taylor expansion of $e^{i T^-}$ is proportional to
\begin{equation}
    \left( T^- \right)^n = 
    \begin{cases}
        \begin{pmatrix}
        |c|^n & 0 \\ 0 & |c|^n
        \end{pmatrix}
        & \quad \text{for $n=$ even,} \\
        \begin{pmatrix}
        0 & c \\ c^* & 0 
        \end{pmatrix}
        \begin{pmatrix}
        |c|^{n-1} & 0 \\ 0 & |c|^{n-1}
        \end{pmatrix}
        & \quad \text{for $n=$ odd.}
    \end{cases}
\end{equation}
The exponential factor is written as
\begin{equation} \label{2exp}
    e^{i T^-} = \cos{|c|} 
    \begin{pmatrix}
    1 & 0 \\ 0 & 1
    \end{pmatrix}
    + \frac{i}{|c|} \sin{|c|}
    \begin{pmatrix}
    0 & c \\ c^* & 0
    \end{pmatrix}.
\end{equation}
Since (\ref{2exp}) must be diagonal, the parameter $|c|$ is constrained to be $|c|=m\pi$ $(m=0,1,2\ldots)$.
Thus, we get the following diagonal matrix:
\begin{equation} \label{diag_2e}
    \text{diag} ( e^{i T^-} ) =
    \begin{cases}
    I_2  & \quad \text{for $m=$ even,} \\
    -I_2 & \quad \text{for $m=$ odd,}
    \end{cases}
\end{equation}
where $I_2$ is an identity matrix.
We find that a diagonal set $[p,q,r,s]$ is not transformed to another diagonal set in both cases, even when $m=$ odd.

\subsection{$N\times N$ matrices with two blocks} \label{sec3.2}
Next we consider the following $N\times N$ generator with two blocks:
\begin{equation} \label{T_2blo}
    T^- =
    \begin{pmatrix}
    0 & A \\
    A^{\dag} & 0 
    \end{pmatrix},
\end{equation}
where $A$ is an $(a \times b)$ sub-matrix $(a,b \geq 2)$ and $0$ is a null sub-matrix. 
(\ref{T_2blo}) is the mixed or anti-commutative type generator in the case that one or more blocks are degenerate, such as $[p,q,r,s=0]$.
The $n$-th term of $e^{i T^-}$ is proportional to
\begin{equation} \label{nth}
    \left( T^- \right)^n = 
    \begin{cases}
        \begin{pmatrix}
        \left(AA^{\dag}\right)^{\frac{n}{2}} & 0 \\
        0 & \left(A^{\dag}A\right)^{\frac{n}{2}}
        \end{pmatrix}
        & \quad \text{for $n=$ even,} \\
        \begin{pmatrix}
        0 & A \\ A^{\dag} & 0 
        \end{pmatrix}
        \begin{pmatrix}
        \left(AA^{\dag}\right)^{\frac{n-1}{2}} & 0 \\
        0 & \left(A^{\dag}A\right)^{\frac{n-1}{2}}
        \end{pmatrix}
        & \quad \text{for $n=$ odd.}
    \end{cases}
\end{equation}
In order to get a diagonal exponential factor, let us transform $T^-$ by unitary matrix: $U = \text{diag} (U_1,U_2)$, 
where $U_1$ and $U_2$ are $(a\times a)$ and $(b\times b)$ unitary sub-matrices.
We note that $P_0$ and $P_1$ are invariant under this unitary transformation.
The $n$-th term (\ref{nth}) is transformed as
\begin{equation} 
    U \left( T^- \right)^n U^{\dag} = 
    \begin{cases}
        \begin{pmatrix}
        \left(A_1\right)^{\frac{n}{2}} & 0 \\
        0 & \left(A_2\right)^{\frac{n}{2}}
        \end{pmatrix}
        & \quad \text{for $n=$ even,} \\
        \begin{pmatrix}
        0 & A' \\ {A'}^{\dag} & 0 
        \end{pmatrix}
        \begin{pmatrix}
        \left(A_1\right)^{\frac{n-1}{2}} & 0 \\
        0 & \left(A_2\right)^{\frac{n-1}{2}}
        \end{pmatrix}
        & \quad \text{for $n=$ odd,}
    \end{cases}
\end{equation}
where we define $A_1 \equiv U_1 (A A^{\dag}) U_1^{\dag}$, $A_2 \equiv U_2 (A^{\dag} A) U_2^{\dag}$, and $A' \equiv U_1 A U_2^{\dag}$.
The two Hermitian matrices $(AA^{\dag})$ and $(A^{\dag}A)$ have the same eigenvalues except for zero.
Following the similar calculations in Section \ref{sec3.1},
the following diagonal exponential factor is obtained:
\begin{equation} \label{diag_N2e}
    U  e^{i T^-} U^\dag =
    \begin{pmatrix}
    \tilde{I}_{a,X} & 0 \\
    0 & \tilde{I}_{b,X}
    \end{pmatrix},
\end{equation}
where $\tilde{I}_{c,X}$ $(c=a,b)$ is a diagonal $(c\times c)$ matrix whose components are $\pm 1$.
The $X$ shows the number of $-1$ $(0 \leq X \leq \min{(a,b)})$.
The rotation by (\ref{diag_N2e}) seemingly moves a diagonal set to another diagonal set, but it cannot be achieved.

\subsection{$N\times N$ matrices with four blocks} \label{sec3.3}
We consider the most general case as the final step:
\begin{equation} \label{T_4blo}
    T^- =
    \begin{pmatrix}
    0 & 0 & 0 & A \\
    0 & 0 & B & 0 \\
    0 & B^{\dag} & 0 & 0 \\
    A^{\dag} & 0 & 0 & 0
    \end{pmatrix},
\end{equation}
where $A$ is a $(p\times s)$ sub-matrix and $B$ is a $(q\times r)$sub-matrix.
The form of (\ref{T_4blo}) is the case of the anti-commutative type (\ref{gene_anti}), but can also be rearranged into the form of the mixed type (\ref{gene_mix}).
From the similar calculation in Section \ref{sec3.2}, the diagonal exponential factor is written as,
\begin{equation} \label{diag_N4e}
    U e^{i T^-} U^\dag =
    \begin{pmatrix}
    \tilde{I}_{p,X} & 0 & 0 & 0 \\
    0 & \tilde{I}_{q,Y} & 0 & 0 \\
    0 & 0 & \tilde{I}_{r,Y} & 0 \\
    0 & 0 & 0 & \tilde{I}_{s,X}
    \end{pmatrix},
\end{equation}
where $X$ and $Y$ denote the numbers of $-1$.
THey satisfy $0\leq X\leq \min{(p,s)}$ and $0\leq Y\leq \min{(q,r)}$.
The rotation by (\ref{diag_N4e}) cannot also achieve the connection between the diagonal sets.
Finally, we conclude that the diagonal sets $[p,q,r,s]$ remains invariant and the non-trivial equivalent relations cannot be obtained under global gauge transformations.


\section{Analysis of local gauge transformations} \label{sec4}


Let us consider local gauge transformations, which have $y$-dependent transformation parameters.
In a brane-world model, the BCs can be chosen differently on the branes and the bulk.
We discuss the local gauge transformations separately on the UV-brane $(\hat{y} = 0)$ and the others $( 0 < \hat{y} \leq 1/2 )$ (see Fig.\ref{FigBrane}).

\subsection{Equivalence classes on the UV-brane} \label{sec4.1}
On the UV-brane $(\hat{y} = 0)$, we prove that the anti-commutative type gauge transformations produce the following equivalent relations \cite{HOSOTANI1989233, HABA2003169}:
\begin{align} \label{UVECs}
\begin{split}
    [p,q,r,s] &\sim [p-1,q+1,r+1,s-1] \quad \text{for $p,s \geq 0$,} \\
    &\sim [p+1,q-1,r-1,s+1] \quad \text{for $q,r \geq 0$.}
\end{split}
\end{align}
Also, It is shown that the commutative and mixed type cannot yield any equivalent relation.
We derive them without relying on an explicit form of gauge transformations, so that the classification of ECs on the UV-brane is completed in this section.

Let us transform $P_i$ $(i=0,1)$ into other diagonal matrices $P'_i$ by the generators $T^{a_{\pm}}$ with the parameters $f^{a_{\pm}}(\hat{y})$.
Using the commutative generator, $P_i$ is transformed as
\begin{align} \label{UV+}
\begin{split}
    P'_i &= 
    \exp{\left[ i  f^{a_+}(\hat{y}_i - \hat{y}) T^{a_+} \right]}
    \exp{\left[ -i f^{a_+}(\hat{y}_i + \hat{y}) T^{a_+} \right]}
    P_i \\
    &\to P_i \quad (\hat{y} \to 0),
\end{split}
\end{align}
where $\hat{y}_0=0$ and $\hat{y}_1=1/2$ are the fixed points on $S^1/Z_2$.
The two exponential factors cancel each other because the parameter functions $f^{a_{+}}(\hat{y})$ are continuous at $\hat{y} = \hat{y}_0, \, \hat{y}_1$.
We find that (\ref{UV+}) is just the identity transformation.
By using the anti-commutative generators, $P_i$ is transformed as
\begin{align} \label{UV-}
\begin{split}
    P'_i &= 
    \exp{\left[ -i f^{a_-}(\hat{y}_i - \hat{y}) T^{a_-} \right]}
    \exp{\left[ -i f^{a_-}(\hat{y}_i + \hat{y}) T^{a_-} \right]}
    P_i \\
    &\to e^{-i c^{a_-}_i T^{a_-} } P_i \quad (\hat{y} \to 0),
\end{split}
\end{align}
where $f^{a_-}(\hat{y}_0) = c^{a_-}_0 /2$ and $f^{a_-}(\hat{y}_1) = c^{a_-}_1 /2$ $\,(c^{a_-}_0,c^{a_-}_1 \in \mathbb{R})$.
From (\ref{UV+}) and (\ref{UV-}), we conclude that a diagonal set $[p,q,r,s]$ do not move to a different diagonal set by the commutative and mixed generators, as shown in the global cases.
On the other hand, under the anti-commutative gauge transformations, $P_0$ and $P_1$ are rotated not simultaneously but independently with $(c^{a_-}_0, c^{a_-}_1)$.
We find that it is different from the global case.

As an example, we independently rotate the set $[p, q=0, r=0, s]$.
$P_0$ and $P_1$ are respectively rotated by
\begin{align}
    D_0 \equiv \text{diag} ( e^{i T^{--}_0} ) =
    \begin{pmatrix}
    \tilde{I}_{p,X} & 0 \\
    0 & \tilde{I}_{s,X}
    \end{pmatrix}, \\
    D_1 \equiv \text{diag} ( e^{i T^{--}_1} ) =
    \begin{pmatrix}
    \tilde{I}_{p,Y} & 0 \\
    0 & \tilde{I}_{s,Y}
    \end{pmatrix},
\end{align}
where $X$ and $Y$ satisfy $0\leq X, Y\leq \min{(p,s)}$ and generally have different values.
We count the number of the same and different eigenvalues between $D_0$ and $D_1$.
Let $t$, $u$, and $v$ be respectively the number of the same $+1$, the same $-1$, and the different eigenvalues ($t+u+v=p+s$, $2u+v=2X+2Y$).
Such transformations produce
\begin{align} 
    [p,q,r,s] \sim [p-v,q+v,r+v,s-v].
\end{align}
This is just the equivalence relations (\ref{UVECs}).


\subsection{Equivalence classes on the Bulk and IR-brane} \label{sec4.2}
Next, we discuss the BCs on the bulk and the IR-brane $( 0 < \hat{y} \leq 1/2 )$.
We show that the commutative gauge transformations produce a new equivalent relation and it connects between all diagonal sets. 

\begin{figure}[ht]
\centering\includegraphics{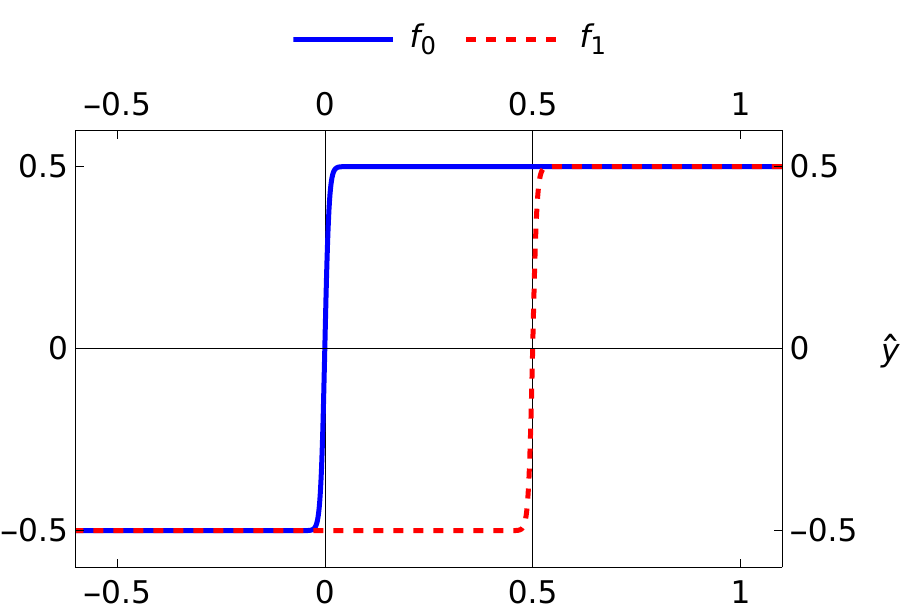}
\caption{The transformation parameters $f_0 (\hat{y})$ and $f_1 (\hat{y})$}
\label{fig_kink}
\end{figure}

We suppose that all of the parameters $\{f^{a_{++}}(\hat{y})\}$ are proportional to a function $f(\hat{y})$, i.e., $f^{a_{++}}(\hat{y})= c^{a_{++}} f(\hat{y})$.
in this case, $P_i$ $(i=0,1)$ are calculated as
\begin{equation} \label{IR+}
    P'_i = \exp{\left[ -i \left\{ f(\hat{y}_i + \hat{y}) - f(\hat{y}_i - \hat{y}) \right\} T^{++} \right]} P_i,
\end{equation}
where $T^{++} \equiv c^{a_{++}} T^{a_{++}}$ is a Hermitian matrix and its form is the commutative type (\ref{gene_com}).
In order to achieve a non-trivial relation, the value of $\{f(\hat{y}_i + \hat{y}) - f(\hat{y_i} - \hat{y})\}$ must be a non-zero constant.
Such a function does not exist on the UV-brane because of the continuity of the parameters, but exists on the bulk and IR-brane.
In order to show that, we consider the following $y$-dependent parameters with a kink at the fixed points $\hat{y} = \hat{y}_0=0$ and $\hat{y} = \hat{y}_1=1/2$ (see Fig.\ref{fig_kink}):
\begin{align}
    f_0 (\hat{y}) &= \frac{1}{2} \tanh{\left[ \lambda \left( \hat{y} - \hat{y}_0 \right) \right]}, \label{UVkink} \\
    f_1 (\hat{y}) &= \frac{1}{2} \tanh{\left[ \lambda \left( \hat{y} - \hat{y}_1 \right) \right]}. \label{IRkink}
\end{align}
Here we adjust the parameter $\lambda$ to a large value and
redefine the area of the bulk and the IR-brane as $1/\lambda \ll \hat{y} \leq 1/2$.
In this case, (\ref{UVkink}) and (\ref{IRkink}) satisfy
\begin{align}
    f_0(\hat{y}_0 + \hat{y}) - f_0(\hat{y}_0 - \hat{y}) =1, \quad
    f_0(\hat{y}_1 + \hat{y}) - f_0(\hat{y}_1 - \hat{y}) =0, \\
    f_1(\hat{y}_0 + \hat{y}) - f_1(\hat{y}_0 - \hat{y}) =0, \quad
    f_1(\hat{y}_1 + \hat{y}) - f_1(\hat{y}_1 - \hat{y}) =1.
\end{align}
Therefore, we find that (\ref{UVkink}) rotates $P_0$ with $P_1$ fixed: $(P'_0, P'_1) = (e^{-i T_0^{++}} \! P_0, \, P_1)$.
(\ref{IRkink}) rotates $P_0$ with $P_1$ fixed: $(P'_0, P'_1) = (P_0, \, e^{-i T_1^{++}} \! P_1)$ under the commutative type gauge transformations.
Here $T_0^{++}$ and $T_1^{++}$ are the generators with $f_0 (\hat{y})$ and $f_1 (\hat{y})$.

Now let us take all sub-matrices in $T^{++}$ diagonal, i.e. $T^{++} = \text{diag}( a_1, a_2, \ldots , a_N )$. 
The exponential factor is written as
\begin{equation}
    e^{i T^{++}} =
    \begin{pmatrix}
    e^{ia_1} & & 0 \\
    & \ddots & \\
    0 & & e^{ia_N}
    \end{pmatrix}.
\end{equation}
Since the transformed matrix $P'$ must be Hermite, the parameters are constrained to $a_i = n_i \pi$ $(i=1,2,\ldots,N \, ; n_i \in \mathbb{Z})$.
Thus, the transformation of $P$ is written as
\begin{equation}
    P' = \tilde{I} P,
\end{equation}
where $\tilde{I}$ is a diagonal matrix with components $+1$ or $-1$. 
It means that the sign of the components of $P_0$ ($P_1$) can be freely flipped while $P_1$ ($P_0$) remains fixed.
As a result, all of the diagonal sets $[p, q, r, s]$ are connected by the commutative gauge transformations with kink parameters.


\section{Conclusions} \label{sec5}
We have classified equivalence classes (ECs) of the boundary conditions (BCs) in $U(N)$ gauge theories on $S^1/Z_2$ brane-world models, where the BCs can be independently chosen on the branes and the bulk.
The diagonal BCs cannot be connected by the global gauge transformations with constant parameters, but the equivalent relations (\ref{UVECs}) have been produced on the UV-brane by the anti-commutative type gauge transformations with coordinate-dependent parameters.
We have proven that there is no connection except for (\ref{UVECs}) on the UV-brane by using not specific but general transformation parameters.
Also, it has been shown that all the diagonal BCs can be connected on the bulk and the IR-brane by the commutative type gauge transformations with a kink at the fixed points.
Therefore, we conclude that the arbitrariness of BCs is solved on the bulk and the IR-brane in this model.
We expect that this work will be useful for completely solving the arbitrariness problem of BCs in extra-dimensional models on compact space.
The different ECs on the UV-brane could be related by using the fact that all the BCs are connected on the bulk and the IR-brane.

A new problem emerges through this study: 
How should we define the physical observable on the bulk and IR-brane?
The fact that all BCs are connected means that the physical mass spectra of the fields are independent of the BCs.
The physical equivalence between the BCs in one EC has been verified by the combination of BCs and dynamical Wilson line phases based on Hosotani mechanism\cite{HOSOTANI1983309, HOSOTANI1983193, HOSOTANI1989233}.
However, this statement is not valid for the commutative type gauge transformations with a kink.
We hope that the study of how the Hosotani mechanism works in this model might lead to a solution to the arbitrariness problem.
We will continue our work and hope to report on this problem.


\section*{Acknowledgment}
The authors would like to thank Kouki Nakamura and Reishi Maeta for useful discussions. We are indebted to members of our laboratory for encouragements.

\bibliographystyle{unsrt} 
\bibliography{main} 

\begin{thebibliography}{10}

\bibitem{MANTON1979141}
N.S. Manton.
\newblock A new six-dimensional approach to the weinberg-salam model.
\newblock {\em Nuclear Physics B}, 158(1):141--153, 1979.

\bibitem{FAIRLIE197997}
D.B. Fairlie.
\newblock Higgs fields and the determination of the weinberg angle.
\newblock {\em Physics Letters B}, 82(1):97--100, 1979.

\bibitem{DBFairlie_1979}
D~B Fairlie.
\newblock Two consistent calculations of the weinberg angle.
\newblock {\em Journal of Physics G: Nuclear Physics}, 5(4):L55, apr 1979.

\bibitem{10.1143/PTP.103.613}
Yoshiharu Kawamura.
\newblock {Gauge Symmetry Reduction from the Extra Space {$S^1/Z_2$}}.
\newblock {\em Progress of Theoretical Physics}, 103(3):613--619, 03 2000.

\bibitem{10.1143/PTP.105.999}
Yoshiharu Kawamura.
\newblock {Triplet-Doublet Splitting, Proton Stability and an Extra Dimension}.
\newblock {\em Progress of Theoretical Physics}, 105(6):999--1006, 06 2001.

\bibitem{PhysRevD.64.055003}
Lawrence Hall and Yasunori Nomura.
\newblock Gauge unification in higher dimensions.
\newblock {\em Phys. Rev. D}, 64:055003, Aug 2001.

\bibitem{doi:10.1142/S0217732302008988}
Masahiro Kubo, C.~S. Lim, and Hiroyuki Yamashita.
\newblock The hosotani mechanism in bulk gauge theories with an orbifold extra
  space {$S^1/Z_2$}.
\newblock {\em Modern Physics Letters A}, 17(34):2249--2263, 2002.

\bibitem{SCRUCCA2003128}
Claudio~A. Scrucca, Marco Serone, and Luca Silvestrini.
\newblock Electroweak symmetry breaking and fermion masses from extra
  dimensions.
\newblock {\em Nuclear Physics B}, 669(1):128--158, 2003.

\bibitem{PhysRevD.69.055006}
Csaba Cs\'aki, Christophe Grojean, Hitoshi Murayama, Luigi Pilo, and John
  Terning.
\newblock Gauge theories on an interval: Unitarity without a higgs boson.
\newblock {\em Phys. Rev. D}, 69:055006, Mar 2004.

\bibitem{BURDMAN20033}
Gustavo Burdman and Yasunori Nomura.
\newblock Unification of higgs and gauge fields in five dimensions.
\newblock {\em Nuclear Physics B}, 656(1):3--22, 2003.

\bibitem{PhysRevD.70.015010}
Naoyuki Haba, Yutaka Hosotani, Yoshiharu Kawamura, and Toshifumi Yamashita.
\newblock Dynamical symmetry breaking in gauge-higgs unification on an
  orbifold.
\newblock {\em Phys. Rev. D}, 70:015010, Jul 2004.

\bibitem{PhysRevD.67.085012}
Csaba Cs\'aki, Christophe Grojean, and Hitoshi Murayama.
\newblock Standard model higgs boson from higher dimensional gauge fields.
\newblock {\em Phys. Rev. D}, 67:085012, Apr 2003.

\bibitem{AGASHE2005165}
Kaustubh Agashe, Roberto Contino, and Alex Pomarol.
\newblock The minimal composite higgs model.
\newblock {\em Nuclear Physics B}, 719(1):165--187, 2005.

\bibitem{PhysRevD.78.096002}
Y.~Hosotani, K.~Oda, T.~Ohnuma, and Y.~Sakamura.
\newblock Dynamical electroweak symmetry breaking in {$SO(5)\times U(1)$}
  gauge-higgs unification with top and bottom quarks.
\newblock {\em Phys. Rev. D}, 78:096002, Nov 2008.

\bibitem{PhysRevD.79.079902}
Y.~Hosotani, K.~Oda, T~Ohnuma, and Y.~Sakamura.
\newblock Erratum: Dynamical electroweak symmetry breaking in {$SO(5)\times
  U(1)$} gauge-higgs unification with top and bottom quarks [phys. rev. d 78,
  096002 (2008)].
\newblock {\em Phys. Rev. D}, 79:079902, Apr 2009.

\bibitem{10.1093/ptep/ptu146}
Shuichiro Funatsu, Hisaki Hatanaka, Yutaka Hosotani, Yuta Orikasa, and Takuya
  Shimotani.
\newblock {Dark matter in the {$SO(5)\times U(1)$} gauge-Higgs unification}.
\newblock {\em Progress of Theoretical and Experimental Physics}, 2014(11), 11
  2014.
\newblock 113B01.

\bibitem{PhysRevD.104.115018}
Shuichiro Funatsu, Hisaki Hatanaka, Yutaka Hosotani, Yuta Orikasa, and Naoki
  Yamatsu.
\newblock Electroweak and left-right phase transitions in {$SO(5)\times
  U(1)\times SU(3)$} gauge-higgs unification.
\newblock {\em Phys. Rev. D}, 104:115018, Dec 2021.

\bibitem{HOSOTANI2005276}
Yutaka Hosotani, Shusaku Noda, and Kazunori Takenaga.
\newblock Dynamical gauge-higgs unification in the electroweak theory.
\newblock {\em Physics Letters B}, 607(3):276--285, 2005.

\bibitem{PANICO2006186}
Giuliano Panico, Marco Serone, and Andrea Wulzer.
\newblock A model of electroweak symmetry breaking from a fifth dimension.
\newblock {\em Nuclear Physics B}, 739(1):186--207, 2006.

\bibitem{PANICO2007189}
Giuliano Panico, Marco Serone, and Andrea Wulzer.
\newblock Electroweak symmetry breaking and precision tests with a fifth
  dimension.
\newblock {\em Nuclear Physics B}, 762(1):189--211, 2007.

\bibitem{10.1093/ptep/ptz083}
Nobuhito Maru and Yoshiki Yatagai.
\newblock {Fermion mass hierarchy in grand gauge-Higgs unification}.
\newblock {\em Progress of Theoretical and Experimental Physics}, 2019(8), 08
  2019.
\newblock 083B03.

\bibitem{PhysRevD.98.015022}
Yuki Adachi and Nobuhito Maru.
\newblock Revisiting electroweak symmetry breaking and the higgs boson mass in
  gauge-higgs unification.
\newblock {\em Phys. Rev. D}, 98:015022, Jul 2018.

\bibitem{PhysRevD.106.055033}
Nobuhito Maru, Haruki Takahashi, and Yoshiki Yatagai.
\newblock Gauge coupling unification in simplified grand gauge-higgs
  unification.
\newblock {\em Phys. Rev. D}, 106:055033, Sep 2022.

\bibitem{doi:10.1142/5326}
M~Harada, Y~Kikukawa, and K~Yamawaki.
\newblock {\em Strong Coupling Gauge Theories and Effective Field Theories}.
\newblock WORLD SCIENTIFIC, 2003.

\bibitem{Quiros:2003gg}
M.~Quiros.
\newblock {New ideas in symmetry breaking}.
\newblock In {\em {Theoretical Advanced Study Institute in Elementary Particle
  Physics (TASI 2002): Particle Physics and Cosmology: The Quest for Physics
  Beyond the Standard Model(s)}}, pages 549--601, 2 2003.

\bibitem{HABA2003169}
Naoyuki Haba, Masatomi Harada, Yutaka Hosotani, and Yoshiharu Kawamura.
\newblock {Dynamical rearrangement of gauge symmetry on the orbifold
  $S^1/Z_2$}.
\newblock {\em Nuclear Physics B}, 657:169--213, 2003.

\bibitem{HOSOTANI1983309}
Yutaka Hosotani.
\newblock Dynamical mass generation by compact extra dimensions.
\newblock {\em Physics Letters B}, 126(5):309--313, 1983.

\bibitem{HOSOTANI1983193}
Yutaka Hosotani.
\newblock Dynamical gauge symmetry breaking as the casimir effect.
\newblock {\em Physics Letters B}, 129(3):193--197, 1983.

\bibitem{HOSOTANI1989233}
Yutaka Hosotani.
\newblock Dynamics of non-integrable phases and gauge symmetry breaking.
\newblock {\em Annals of Physics}, 190(2):233--253, 1989.

\bibitem{10.1143/PTP.111.265}
Naoyuki Haba, Yutaka Hosotani, and Yoshiharu Kawamura.
\newblock {Classification and Dynamics of Equivalence Classes in SU(N) Gauge
  Theory on the Orbifold $S^1/Z_2$}.
\newblock {\em Progress of Theoretical Physics}, 111(2):265--289, 02 2004.

\bibitem{PhysRevD.69.125014}
Yutaka Hosotani, Shusaku Noda, and Kazunori Takenaga.
\newblock Dynamical gauge symmetry breaking and mass generation on the orbifold
  ${T}^{2}{/Z}_{2}$.
\newblock {\em Phys. Rev. D}, 69:125014, Jun 2004.

\bibitem{10.1143/PTP.120.815}
Yoshiharu Kawamura, Teppei Kinami, and Takashi Miura.
\newblock {Equivalence Classes of Boundary Conditions in Gauge Theory on $Z_3$
  Orbifold}.
\newblock {\em Progress of Theoretical Physics}, 120(5):815--831, 11 2008.

\bibitem{10.1143/PTP.122.847}
Yoshiharu Kawamura and Takashi Miura.
\newblock {Equivalence Classes of Boundary Conditions in SU(N) Gauge Theory on
  2-Dimensional Orbifolds}.
\newblock {\em Progress of Theoretical Physics}, 122(4):847--864, 10 2009.

\bibitem{doi:10.1142/S0217751X20502061}
Yoshiharu Kawamura and Yasunari Nishikawa.
\newblock On diagonal representatives in boundary condition matrices on
  orbifolds.
\newblock {\em International Journal of Modern Physics A}, 35(31):2050206,
  2020.

\bibitem{Kawamura2023}
Yoshiharu Kawamura, Eiji Kodaira, Kentaro Kojima, and Toshifumi Yamashita.
\newblock On representation matrices of boundary conditions in {$SU(n)$} gauge
  theories compactified on two-dimensional orbifolds.
\newblock {\em \textit{Journal of High Energy Physics}}, 04(113), 2023.

\bibitem{10.1093/ptep/ptae027}
Kota Takeuchi and Tomohiro Inagaki.
\newblock {New Classification Method for Equivalence Classes on ${S^1/Z_2}$ and
  ${T^2/Z_3}$ Orbifolds}.
\newblock {\em Progress of Theoretical and Experimental Physics},
  2024(3):033B03, 02 2024.

\end{thebibliography}

\end{document}